\documentclass[draftcls,onecolumn]{IEEEtran}

\usepackage{fixltx2e} 
\ifCLASSINFOpdf
  \usepackage[pdftex]{graphicx}
\else
  \usepackage{graphicx}
\fi
\usepackage{psfrag}
\usepackage[cmex10]{amsmath}
\usepackage{amsfonts,amssymb,latexsym,hyperref}
\usepackage[nospace]{cite}
\usepackage[usenames,dvipsnames]{color}

 \newcommand{\capFrag}[2]{}
 \newcommand{\capTable}[2]{}

 \newcommand{\defn}{\triangleq}

 \newcommand{\hvec}[1]{\ensuremath{\Hat{\boldsymbol{#1}}}}
 
 \renewcommand{\vec}[1]{\ensuremath{\boldsymbol{#1}}}

 \newcommand{\mc}[1]{\ensuremath{\mathcal{#1}}}

 \newcommand{\Real}{{\mathbb{R}}}

 \newcommand{\tran}{^{\text{\textsf{T}}}}

 \renewcommand{\eqref}[1]{(\ref{eq:#1})}

 \newcommand{\figref}[1]{Fig.~\ref{fig:#1}}
 \newcommand{\tabref}[1]{Table~\ref{tab:#1}}

 \newcommand{\true}{_0}

\begin{document}

\title{Denoising-based Vector AMP}

\author{%
\IEEEauthorblockN{
Philip Schniter\IEEEauthorrefmark{1}, 
Sundeep Rangan\IEEEauthorrefmark{2}, 
and Alyson Fletcher\IEEEauthorrefmark{3}}
 
\IEEEauthorblockA{\IEEEauthorrefmark{1}
Department of Electrical and Computer Engineering,
The Ohio State University,
Columbus, OH.}
\IEEEauthorblockA{\IEEEauthorrefmark{2} 
Department of Electrical and Computer Engineering,
New York University, 
Brooklyn, NY.}
\IEEEauthorblockA{\IEEEauthorrefmark{3}
Departments of Statistics, Mathematics, and Electrical Engineering, 
University of California, 
Los Angeles, CA.}
}

\maketitle

\begin{abstract}
The D-AMP methodology, recently proposed by Metzler, Maleki, and Baraniuk, allows one to plug in sophisticated denoisers like BM3D into the AMP algorithm to achieve state-of-the-art compressive image recovery.
But AMP diverges with small deviations from the i.i.d.-Gaussian assumption on the measurement matrix.
Recently, the VAMP algorithm has been proposed to fix this problem. 
In this work, we show that the benefits of VAMP extend to D-VAMP.
\end{abstract}


\vspace*{9pt}

Consider the problem of recovering a (vectorized) image $\vec{x}\true\in\Real^N$ from compressive (i.e., $M\ll N$) noisy linear measurements
\begin{align}
\vec{y}
&= \vec{\Phi x}\true + \vec{w} \in \Real^M
\label{eq:yx} ,
\end{align}
known as ``compressive imaging.''
The ``sparse'' approach to this problem exploits sparsity in the coefficients $\vec{v}\true\defn\vec{\Psi x}\true\in\Real^N$ of an orthonormal wavelet transform $\vec{\Psi}$.
The idea is to rewrite \eqref{yx} as
\begin{align}
\vec{y}
&= \vec{A v}\true + \vec{w} 
\text{~~for~~} \vec{A}\defn \vec{\Phi \Psi}\tran 
\label{eq:yv} ,
\end{align}
recover an estimate $\hvec{v}$ of $\vec{v}\true$ from $\vec{y}$, and then construct the image estimate as $\hvec{x}=\vec{\Psi}\tran\hvec{v}$.

Although many algorithms have been proposed for sparse recovery of $\vec{v}\true$, a notable one is the approximate message passing (AMP) algorithm from \cite{Donoho:PNAS:09}.
It is computationally efficient (i.e., one multiplication by $\vec{A}$ and $\vec{A}\tran$ per iteration and relatively few iterations) and its performance, when 
$M$ and $N$ are large and $\vec{\Phi}$ is zero-mean i.i.d.\ Gaussian, is rigorously characterized by a scalar state evolution. 

A variant called ``denoising-based AMP'' (D-AMP) was recently proposed \cite{Metzler:TIT:16} for \emph{direct} recovery of $\vec{x}\true$ from \eqref{yx}.
It exploits the fact that, at iteration $t$, AMP constructs a pseudo-measurement of the form $\vec{v}\true+\mc{N}(\vec{0},\sigma_t^2\vec{I})$ with known $\sigma_t^2$, which is amenable to any image denoising algorithm.
By plugging in a state-of-the-art image denoiser like BM3D \cite{Dabov:TIP:07}, D-AMP yields state-of-the-art compressive imaging. 

AMP and D-AMP, however, have a serious weakness: they diverge under small deviations from the zero-mean i.i.d.\ Gaussian assumption on $\vec{\Phi}$, such as non-zero mean or mild ill-conditioning.
A robust alternative called ``vector AMP'' (VAMP) was recently proposed \cite{Rangan:VAMP}.
VAMP has similar complexity to AMP and a rigorous state evolution that holds under right-rotationally invariant $\vec{\Phi}$---a much larger class of matrices.
Although VAMP needs to know the variance of the measurement noise $\vec{w}$, an auto-tuning method was proposed in \cite{Fletcher:EMVAMP}.

In this work, we integrate the D-AMP methodology from \cite{Metzler:TIT:16} into auto-tuned VAMP from \cite{Fletcher:EMVAMP}, leading to ``D-VAMP.''
(For a matlab implementation, see \url{http://dsp.rice.edu/software/DAMP-toolbox}.)

\begin{figure}[t]
\centering
\psfrag{BM3D-AMP}[B][B][0.8]{\textsf{\textbf{BM3D-AMP}}}
\psfrag{BM3D-GEC}[B][B][0.8]{\textsf{\textbf{BM3D-VAMP}}}
\psfrag{PSNR}[B][B][0.7]{\textsf{PSNR}}
\psfrag{iteration}[t][t][0.7]{\textsf{iteration}}
\includegraphics[width=0.8\columnwidth,trim=0 10 0 0,clip]{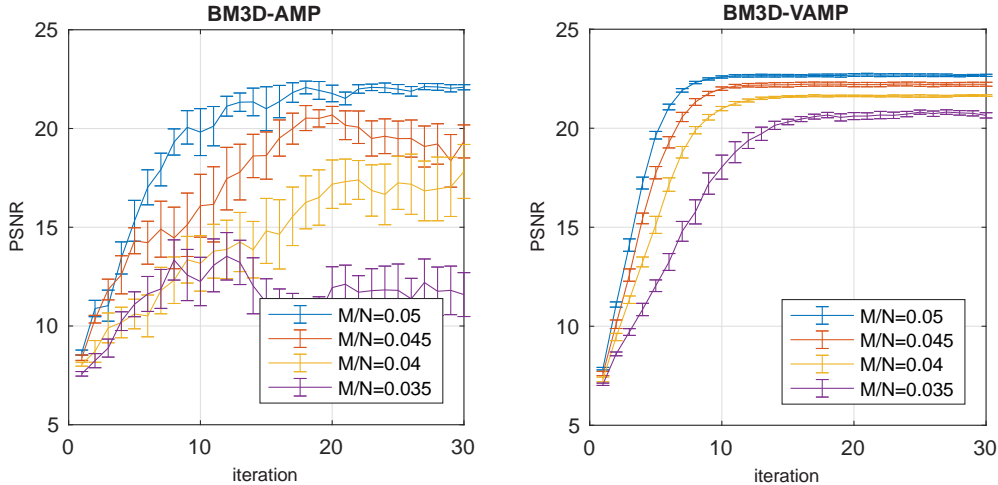}
\caption{PSNR versus iteration at several sampling ratios $M/N$ for i.i.d.\ Gaussian $\vec{A}$.}
\label{fig:dvamp_lowrate}
\end{figure}

To test D-VAMP, we recovered the $128\times 128$ \textsl{lena, barbara, boat, fingerprint, house}, and \textsl{peppers} images using $10$ realizations of $\vec{\Phi}$.
\tabref{vsRate} shows that, for i.i.d.\ Gaussian $\vec{\Phi}$, the average PSNR and runtime of D-VAMP is similar to D-AMP at medium sampling ratios.
The PSNRs for $\vec{v}$-based indirect recovery, using Lasso (i.e., ``$\ell_1$'')-based AMP and VAMP, are significantly worse.
At small sampling ratios, D-VAMP behaves better than D-AMP, as shown in \figref{dvamp_lowrate}.

To test robustness to ill-conditioning in $\vec{\Phi}$, we constructed $\vec{\Phi}=\vec{JSPFD}$, with $\vec{D}$ a diagonal matrix of random $\pm 1$, $\vec{F}$ a (fast) Hadamard matrix, $\vec{P}$ a random permutation matrix, and $\vec{S}\in\Real^{M\times N}$ a diagonal matrix of singular values.
The sampling rate was fixed at $M/N=0.1$,
the noise variance chosen to achieve SNR=$32$~dB,
and the singular values were geometric, i.e., $s_i/s_{i-1}=\rho~\forall i>1$, with $\rho$ chosen to yield a desired condition number.
\tabref{vsCondNum} shows that (D-)AMP breaks when the condition number is $\geq 10$, whereas (D-)VAMP shows only mild degradation in PSNR (but not runtime).

\begin{table}[h]
\caption{Average PSNR and runtime from measurements with i.i.d.\ Gaussian matrices and zero noise after $30$ iterations}
\label{tab:vsRate}
\centering
\begin{tabular}{@{}c@{~}|| @{\,}c@{~}c@{\,}|@{\,}c@{~}c@{\,}|@{\,}c@{~}c@{\,}|@{\,}c@{~}c@{\,}|@{\,}c@{~}c@{}}
sampling ratio
& \multicolumn{2}{@{\,}c@{\,}|@{\,}}{10\%} 
& \multicolumn{2}{@{\,}c@{\,}|@{\,}}{20\%} 
& \multicolumn{2}{@{\,}c@{\,}|@{\,}}{30\%} 
& \multicolumn{2}{@{\,}c@{\,}|@{\,}}{40\%} 
& \multicolumn{2}{@{\,}c@{}}{50\%} \\\hline\hline
& PSNR & time & PSNR & time & PSNR & time & PSNR & time & PSNR & time\\\hline
$\ell_1$-AMP & \bf17.7&\bf0.5s & \bf20.2&1.0s & \bf22.4&1.6s & 24.6&2.3s & 27.0&3.1s\\
$\ell_1$-VAMP & 17.6&\bf0.5s & \bf20.2&\bf0.9s & \bf22.4&\bf1.4s & \bf24.8&\bf1.8s & \bf27.2&\bf2.3s\\\hline
BM3D-AMP & \bf25.2&\bf10.1s& \bf30.0&8.8s & \bf32.5&8.6s & 35.1&9.1s & 37.4&9.8s\\
BM3D-VAMP & \bf25.2&10.4s & \bf30.0&\bf8.5s & \bf32.5&\bf8.2s & \bf35.2&\bf8.5s & \bf37.7&\bf8.8s\\
\end{tabular}
\end{table}

\begin{table}[h]
\vspace{-2mm}
\caption{Average PSNR and runtime from measurements with DHT-based matrices and SNR=32~dB after $10$ iterations}
\label{tab:vsCondNum}
\centering
\begin{tabular}{@{}c@{~}|| @{\,}c@{~}c@{\,}|@{\,}c@{~}c@{\,}|@{\,}c@{~}c@{\,}|@{\,}c@{~}c@{\,}|@{\,}c@{~}c@{}}
condition no.
& \multicolumn{2}{@{\,}c@{\,}|@{\,}}{1} 
& \multicolumn{2}{@{\,}c@{\,}|@{\,}}{10} 
& \multicolumn{2}{@{\,}c@{\,}|@{\,}}{10$^2$} 
& \multicolumn{2}{@{\,}c@{\,}|@{\,}}{10$^3$} 
& \multicolumn{2}{@{\,}c@{}}{10$^4$} \\\hline\hline
& PSNR & time & PSNR & time & PSNR & time & PSNR & time & PSNR & time\\\hline
$\ell_1$-AMP & 17.3&\bf0.02 & $<$0&--- & $<$0&--- & $<$0&--- & $<$0&---\\
$\ell_1$-VAMP & \bf17.4&0.04 & \bf17.4&\bf0.04 & \bf15.6&\bf0.03 & \bf14.7&\bf0.03 & \bf14.4&\bf0.03\\\hline
BM3D-AMP & \bf24.8&\bf5.2s & 8.0&--- & 7.2&--- & 7.1&--- & 7.2&---\\
BM3D-VAMP & \bf24.8&5.4s & \bf24.3&\bf5.5s & \bf22.6&\bf5.3s & \bf21.4&\bf4.9s & \bf20&\bf4.5s\\
\end{tabular}
\end{table}

\bibliographystyle{IEEEtran}
\bibliography{macros_abbrev,sparse,machine}

\end{document}